%Paper: hep-ph/9402204
%From: Koichi FUNAKUBO <funakubo@himiko.cc.saga-u.ac.jp>
%Date: Wed, 02 Feb 94 11:15:45 +0900
%Date (revised): Sat, 05 Feb 94 18:41:34 +0900

% ---------------------------------------------------------------
%    Fermion Scattering off CP-violating Electroweak Bubble Wall
%
%       required macros : phyzzx, tables and epsf
%    Uuencoded tar.Z file of two PostScript files is appended
%    at the end of the text.
% ----------------------------------------------------------------
\input phyzzx
\input tables
%%
%%--- local extension
%%
\catcode`\@=11
\newfam\mibfam
% Extra fonts
%
\font\sevensl =cmsl8 at 7pt
\font\sevenit =cmti7
\font\seventeenmib =cmmib10 scaled\magstep3 \skewchar\seventeenmib='177
\font\fourteenmib =cmmib10 scaled\magstep2 \skewchar\fourteenmib='177
\font\twelvemib =cmmib10 scaled\magstep1 \skewchar\twelvemib='177
\font\tenmib =cmmib10 \skewchar\tenmib='177
\font\ninemib =cmmib9 \skewchar\ninemib='177
\font\sixmib =cmmib6 \skewchar\sixmib='177
%
% Redefinition of fourteenf@nts
\def\fourteenf@nts{\relax
\textfont0=\fourteenrm \scriptfont0=\tenrm
\scriptscriptfont0=\sevenrm
\textfont1=\fourteeni \scriptfont1=\teni
\scriptscriptfont1=\seveni
\textfont2=\fourteensy \scriptfont2=\tensy
\scriptscriptfont2=\sevensy
\textfont3=\fourteenex \scriptfont3=\twelveex
\scriptscriptfont3=\tenex
\textfont\itfam=\fourteenit \scriptfont\itfam=\tenit
\textfont\slfam=\fourteensl \scriptfont\slfam=\tensl
\textfont\bffam=\fourteenbf \scriptfont\bffam=\tenbf
\scriptscriptfont\bffam=\sevenbf
\textfont\mibfam=\fourteenmib \scriptfont\mibfam=\tenmib
\textfont\ttfam=\fourteentt
\textfont\cpfam=\fourteencp }
% Redefinition of twelvef@nts
\def\twelvef@nts{\relax
\textfont0=\twelverm \scriptfont0=\ninerm
\scriptscriptfont0=\sixrm
\textfont1=\twelvei \scriptfont1=\ninei
\scriptscriptfont1=\sixi
\textfont2=\twelvesy \scriptfont2=\ninesy
\scriptscriptfont2=\sixsy
\textfont3=\twelveex \scriptfont3=\tenex
\scriptscriptfont3=\tenex
\textfont\itfam=\twelveit \scriptfont\itfam=\nineit
\textfont\slfam=\twelvesl \scriptfont\slfam=\ninesl
\textfont\bffam=\twelvebf \scriptfont\bffam=\ninebf
\scriptscriptfont\bffam=\sixbf
\textfont\mibfam=\twelvemib \scriptfont\mibfam=\ninemib
\scriptscriptfont\mibfam=\sixmib
\textfont\ttfam=\twelvett
\textfont\cpfam=\twelvecp }
% Redefintion of tenf@nts
\def\tenf@nts{\relax
\textfont0=\tenrm \scriptfont0=\sevenrm
\scriptscriptfont0=\fiverm
\textfont1=\teni \scriptfont1=\seveni
\scriptscriptfont1=\fivei
\textfont2=\tensy \scriptfont2=\sevensy
\scriptscriptfont2=\fivesy
\textfont3=\tenex \scriptfont3=\tenex
\scriptscriptfont3=\tenex
\textfont\itfam=\tenit \scriptfont\itfam=\sevenit
\textfont\slfam=\tensl \scriptfont\slfam=\sevensl
\textfont\bffam=\tenbf \scriptfont\bffam=\sevenbf
\scriptscriptfont\bffam=\fivebf
\textfont\mibfam=\tenmib
\textfont\ttfam=\tentt
\textfont\cpfam=\tencp }
\def\mib{\n@expand\f@m\mibfam}
\Twelvepoint
\catcode`\@=12
%%
%%--- end of local extension
%%
\tolerance=9999
\overfullrule=0pt
\def\e{{\rm e}}
\def\del{\partial}
\def\dslash{\del\kern-0.55em\raise 0.14ex\hbox{/}}
\def\Lag{{\cal L}}

\def\bfx{{\mib x}}
\def\bfpT{{{\mib p}_T}}
\def\abs#1{{\left|{#1}\right|}}
\def\expecv#1{\langle #1 \rangle}
\def\phipa{\phi^{(+\alpha)}}
\def\phima{\phi^{(-\alpha)}}
\def\phipsa{\phi^{(+\sigma\alpha)}}
\def\phimsa{\phi^{(-\sigma\alpha)}}
\def\a{\alpha}
\def\b{\beta}
\def\ep{\epsilon}
\def\g{\gamma}
\def\G{\Gamma}
\def\s{\sigma}
\def\t{\tau}
\def\scien#1#2{${#1}\times 10^{#2}$}
\date={January 29, 1994}
\Pubnum={SAGA--HE--55  \cr
         KYUSHU--HET--15}
\Pubtype={}
\titlepage
\title{Fermion Scattering off CP-Violating Electroweak Bubble Wall}
\bigskip\bigskip
\centerline{Koichi Funakubo${}^{1)}$\footnote{\#a}
{e-mail: \ funakubo@sagagw.cc.saga-u.ac.jp},\ \ \  Akira Kakuto${}^{2)}$%
\footnote{\#b}{e-mail: \ kakuto@fuk.kindai.ac.jp},
\ \ Shoichiro Otsuki${}^{2)}$,}
\centerline{Kazunori Takenaga${}^{3)}$\footnote{\#c}%
{e-mail: \ f77498a@kyu-cc.cc.kyushu-u.ac.jp}
\ \ and\ \ \  Fumihiko Toyoda${}^{2)}$\footnote{\#d}%
{e-mail: \ ftoyoda@fuk.kindai.ac.jp}}
\bigskip
\centerline{${}^{1)}$\it Department of Physics, Saga University,
Saga 840 Japan}
\centerline{${}^{2)}$\it Department of Liberal Arts, Kinki University in
Kyushu, Iizuka 820 Japan}
\centerline{${}^{3)}$\it Department of Physics, Kyushu University,
Fukuoka 812 Japan}
\bigskip\bigskip
%%%%%%%%%
\abstract{
A general prescription to solve the Dirac equation in the presence of
CP-violating electroweak bubble wall is presented.
The profile of the bubble wall is not specified except that the
wall height is $m_0$ and zero deep in the broken- and the
symmetric-phase regions, respectively, where $m_0$ is a fermion
mass given by the Higgs-vacuum-expectation value and the Yukawa coupling.
The CP-violating effects are evaluated by regarding CP-violating
part of the bubble wall as a perturbation to
CP-conserving solutions. 
The basic quantity, $R_{R\rightarrow L}-\bar R_{R\rightarrow L}$,
which would contribute to the cosmological
baryon asymmetry, is estimated for some typical profiles of the wall,
where $R_{R\rightarrow L}$($\bar R_{R\rightarrow L}$) is the reflection
coefficient of right-handed chiral fermion (anti-fermion).
}
\vfill\eject
%
%
%
%%%%%%%%%%%%
% references
%%%%%%%%%%%%
\REF\SAKH{A.~Sakharov, JETP Lett. {\bf 5} (1967) 24.}
\REF\KRS{G.~'t Hooft, Phys.~Rev.~Lett. {\bf 37} (1976) 8;\nextline
V.~Kuzmin, V.~Rubakov and M.~Shaposhnikov,
Phys.~Lett. {\bf B155} (1985) 36.}
\REF\BRA{D.~E.~Brahm, Proceedings of the XXVI International
Conference on High Energy Physics (1993) 1583, and references therein.}
\REF\KOM{M.~Kobayashi and T.~Maskawa,
Prog.~Theor.~Phys. {\bf 49} (1973) 652.}
\REF\TZ{N.~Turok and J.~Zadrozny,
Nucl.~Phys. {\bf B369} (1992) 729.}
\REF\FKT{K.~Funakubo, A.~Kakuto and K.~Takenaga, \nextline
preprint KYUSHU--HET--10---SAGA--HE--50 (1993), to be published in
Prog.~Theor.~Phys. {\bf 91} No.~2 (1994).}
\REF\CKN{A.~Cohen, D.~Kaplan and A.~Nelson,
Nucl.~Phys. {\bf B373} (1992) 453.}
\REF\AJMV{A.~Ayala, J.~Jalilian-Marian, L.~McLerran and
A.~P.~Vischer, \ \ preprint \nextline
TPI--MINN--54---MUC--MINN--93/30--T---UMN--TH--1226/93 (1993).}
\REF\CH{See for example, R.~Courant and D.~Hilbert,\nextline
Methods of Mathematical Physics ( Interscience, New York ) (1953).}
\REF\HT{M.~Fukugita and T.~Yanagida,\nextline
Phys.~Lett. {\bf B174} (1986) 45; Phys.~Rev. {\bf D42} (1990) 1285;\nextline
J.~A.~Harvey and M.~S.~Turner, Phys.~Rev. {\bf D42} (1990) 3344.}
%%%%%%%%%%%%%%%%%%%
% end of references
%%%%%%%%%%%%%%%%%%%
%
%
%
\chapter{Introduction}
It is one of the most challenging problems in particle physics and
cosmology to explain the baryon asymmetry of the universe.
In order to generate the baryon asymmetry, the basic theory
should at least satisfy the famous three conditions\refmark\SAKH\ :
there exist (i) baryon-number violation, (ii) CP nonconservation
and (iii) out-of-equilibrium processes. It has been recognized that
the baryon number is violated in the standard model through change of
gauge-field configurations due to the axial anomaly
\refmark\KRS. At very high temperature, the baryon number is
badly broken even to erase any primordial asymmetry, while the transition
rate becomes negligible at low temperature. Thus there is a potential
possibility in the standard model to produce the baryon asymmetry
if the electroweak phase transition at a temperature
$T\sim 100$ GeV is of first order, which guarantees the above condition
(iii), and if CP-nonconserving interactions are effective.
It is known that the phase transition in the standard model may be
weakly of first order for light Higgs boson.  However, the experimental
lower bound on the Higgs-boson mass ($\sim 60$ GeV) seems to contradict
the requirement to forbid anomalous processes induced by the
sphaleron after the electroweak phase transition\refmark\BRA.
Furthermore it seems that the CP violation in the
Kobayashi-Maskawa scheme\refmark\KOM\ is too feeble to generate
the observed baryon asymmetry.
\par
A simple generalization of the standard model would be
to introduce another Higgs doublet\refmark\TZ.
In two-Higgs-doublet models,
including the supersymmetric extension of the standard model,
there could appear a new CP-violating
phase in the Higgs sector.
It is expected that the new CP violation could make baryon
asymmetry large enough to be comparable with the observed value.
Moreover it is argued that the constraint on the mass of the lightest
Higgs boson to avoid washing out of the baryon asymmetry after
the electroweak phase transition is not so stringent that the Higgs
boson could be as heavy as about 100GeV \refmark\TZ.
It is quite possible that the phase transition may be of first order
because of many degrees of freedom in the Higgs sector\refmark\TZ.
In a previous paper, three of the authors
( K.~F., A.~K. and K.~T. ) have shown
that the electroweak phase transition is clearly of first order in
a two-Higgs-doublet model with a constraint that the
Higgs potential contains no quadratic terms in the symmetric
phase\refmark\FKT.
\par
In this paper, we study fermion propagation in CP-violating
bubble-wall background, which would play an important
role to generate the cosmological baryon asymmetry\refmark\CKN.
We give a general prescription to solve the Dirac equation in the
presence of CP-violating bubble wall.\foot{For a special bubble-wall 
profile of the
kink type with no CP violation, the Dirac equation has been studied by
Ayala {\it et~al.\/\/} \ \refmark\AJMV.}
For simplicity, we take a single species of fermion,
{\it e.g.,}\/ the top quark.
If the electroweak phase transition is actually of first order,
supercooling will occur and bubbles of the broken phase will be generated
in false vacuum.  The bubbles will expand with some velocity and
the space will be filled with the true vacuum.
Since the bubble scale is macroscopic,
the profile function of the bubble wall may be assumed to depend on
a single spatial coordinate in the rest frame of the bubble wall.
Far from the wall, deep in the broken-phase region, the Higgs field
develops a constant vacuum expectation value. We assign a complex
mass to the fermion as a function of $z$, $m(z) = m_R(z) + im_I(z)$,
where $z$ is the coordinate perpendicular to the wall.
The real part $m_R(z)$ asymptotically behaves such that
$m_R(z)\rightarrow 0$ as $z\rightarrow -\infty$ ( symmetric phase )
and $m_R(z)\rightarrow m_0$ as $z\rightarrow +\infty$ ( broken phase ),
where $m_0$ is the fermion mass. The imaginary part $m_I(z)$
gives rise to CP violation if $m_I(z) / m_R(z)$ is not a constant.
\par
We do not specify any explicit form of the profile function
$m(z)$ and obtain general expressions for the reflection and the
transmission coefficients.
The CP-violating effects are evaluated by regarding the CP-violating term
as a perturbation to CP-conserving solutions.
Namely, we first solve the CP-conserving Dirac equation.
Subsequently we solve the full equation to the first order of the
CP-violating term by means of the Green's function\refmark\CH.
Our method corresponds to the distorted wave Born approximation
( DWBA ). The reflection coefficient $R$ and the transmission
coefficient $T$ for the fermion incident from the symmetric phase region
to the wall satisfy the unitarity $R+T=1$ to the first order
of perturbation theory.
The basic quantity, $\Delta R\equiv R_{R\rightarrow L} - \bar
R_{R\rightarrow L}$,
which would be important to explain the cosmological
baryon asymmetry\refmark\CKN, is estimated for some typical profiles
of the wall, where $R_{R\rightarrow L}$ ( $\bar R_{R\rightarrow L}$ )
is the reflection coefficient for right-handed chiral fermion
(anti-fermion).
The absolute value of $\Delta R$ is found to decrease rapidly with 
the increase of $E^* \equiv \sqrt{E^2 - {\mib p}_T^2}$, 
where $E$ and ${\mib p}_T$ are energy and transverse momentum of 
the incident fermion, respectively. 
It seems that $\abs{\Delta R}$ decreases more rapidly with the increase
of $E^*$ for thicker walls.
We expect that our prescription would serve to construct new models
to generate the cosmological baryon asymmetry in the electroweak
phase transition of first order.
\par
In the next section, we will present a general prescription
to solve the Dirac equation in the background of CP-violating bubble wall.
The reflection and the transmission coefficients are obtained from the
solutions. In Section 3, the results in Section 2 are applied to the
special case where the real part of the bubble-wall profile is of
the kink type.\refmark\AJMV\  
In this case we estimate $\Delta R$
numerically by assuming typical forms for the imaginary part of the
bubble-wall profile. The final section is devoted to conclusions.
Details of calculations will be given in Appendix.
%
% end of Introduction
%
%
\chapter{DWBA to CP-Violating Dirac Equation}
\noindent
\section{Dirac equation and Ansatz}
Here we solve the Dirac equation in the bubble-wall background
with CP violation. For simplicity, we consider one-flavor model
described by the lagrangian,
$$
  \Lag = \bar\psi_L i\dslash \psi_L + \bar\psi_R i\dslash \psi_R
         + ( f \bar\psi_L\psi_R\phi + {\rm h.c.} ).
\eqn\IIa$$
In the vacuum, near the first-order phase transition point,
$\expecv\phi$ may be $x$-dependent field, so that we put
$$
  m(\bfx) = - f \expecv\phi(\bfx),
\eqn\IIb$$
where $m(\bfx)$ is in general complex-valued and we neglect
the time dependence. If the phase of $m(\bfx)$ has no spatial
dependence, it is removed by a constant bi-unitary transformation.
We are interested in \bfx-dependent phase which cannot be transformed
away.
The Dirac equation to be solved is
$$
i\dslash\psi(t,\bfx)-m(\bfx)P_R\psi(t,\bfx)-m^*(\bfx)P_L\psi(t,\bfx) = 0,
\eqn\IIc$$
where $P_R = (1+\gamma_5)/2$ and $P_L = (1-\gamma_5)/2$.
In the bubble-wall background whose radius is sufficiently large, 
$m(\bfx)$ can be regarded as
a function of only one spatial coordinate, so that we put
$m(\bfx)=m(z)$.\par
To solve \IIc, we take the following Ansatz:
$$\eqalign{
 \psi(t,\bfx) 
 &= (i\dslash + m^*(z)P_R + m(z)P_L) 
    \e^{i\s(-Et+\bfpT\bfx_T)}\psi_E(\bfpT,z) \cr
 &= \e^{i\s(-Et+\bfpT\bfx_T)}[\s(\g^0E-\g_Tp_T)
    + i\g^3\del_z+ m^*(z)P_R + m(z)P_L] \psi_E(\bfpT,z), \cr
}\eqn\IId$$
where $\s=+(-)$ for positive(negative)-energy solution,
$\bfpT=(p^1,p^2)$, $\bfx_T=(x^1,x^2)$, $p_T=\abs{\bfpT}$ and
$\g_Tp_T=\g^1p^1+\g^2p^2$. If we put $E=E^*\cosh\eta$ and
$p_T=E^*\sinh\eta$ with $E^*=\sqrt{E^2-p_T^2}$, the Lorentz
transformation of the spinor, represented by 
$\psi^\prime=\e^{\eta\g^0\g_T/2}\psi$, eliminates $\bfpT$ from
\IId\  and from the Dirac equation. Suppose that we do this Lorentz
rotation for a fixed $\bfpT$.
Then the Dirac equation is rewritten, in the second-order form, as
$$
 \bigl[ {E^*}^2 + \del_z^2-{\abs{m(z)}}^2
    +im_R^\prime(z)\g^3 - m_I^\prime(z)\g_5\g^3 \bigr]\psi_E(z)=0,
\eqn\IIe$$
where $m(z)=m_R(z)+im_I(z)$, with both $m_R(z)$ and $m_I(z)$
being real-valued. Now let us introduce a set of dimensionless 
variables,
using a parameter, $a$, which has a dimension of mass and whose
inverse characterizes the thickness of the wall. Define
$$\eqalign{
 & m_R(z)=m_0f(az)=m_0f(x), \cr
 & m_I(z)=m_0g(az)=m_0g(x), \cr
 & x\equiv az, \qquad \t\equiv at,\qquad
   \ep \equiv E^*/a,\qquad \xi\equiv m_0/a,  \cr
}\eqn\IIf$$
where $m_0$ is the fermion mass in the broken phase.
Then the Ansatz \IId\  and the second-order equation \IIe\  are
expressed as
$$
  \psi(\t,x) = \e^{-i\s\ep\t}[\s\ep\g^0 + 
  i\g^3{d\over{dx}}+\xi f(x) - i\xi g(x)\g_5]\psi_\ep(x),
\eqn\IIg$$
and
$$
 [\ep^2 + {d^2\over{dx^2}} - \xi^2(f(x)^2+g(x)^2)
  +i\xi f^\prime(x)\g^3-\xi g^\prime(x)\g_5\g^3]
  \psi_\ep(x) = 0,
\eqn\IIh$$
respectively.\par
Our aim is to solve \IIh\  and put it back into the Ansatz
\IIg\  to obtain the Dirac spinor, and to compute various
currents from which the reflection and transmission 
coefficients are defined.
As for $f(x)$ and $g(x)$, we do not specify their functional
forms but only assume that
$$
 f(x)\rightarrow\cases{1, &as $x\rightarrow+\infty$,\cr
                       0, &as $x\rightarrow-\infty$,\cr}
\eqn\IIi$$
and that $\abs{g(x)}<<1$, which means that the CP violation is small.
\IIi\  means that the system is in the broken phase at $x\sim+\infty$
while in the symmetric phase at $x\sim-\infty$.
\par
\section{Outline of the DWBA to the Dirac equation}
Let us explain our strategy to construct the spinor solution.
Since $\abs{g(x)}$ is very small, we regard it as a perturbation
and keep only quantities of $O(g^1)$. First, put
$$
   \psi_\ep(x) = \psi^{(0)}(x) + \psi^{(1)}(x),
\eqn\IIj$$
where $\psi^{(0)}(x)$ is a solution to the unperturbed equation
$$
 [\ep^2 + {d^2\over{dx^2}} - \xi^2f(x)^2+i\xi f^\prime(x)\g^3]
  \psi^{(0)}(x) = 0,
\eqn\IIk$$
with an appropriate boundary condition.
Then $\psi^{(1)}(x)$ can be solved as
$$
 \psi^{(1)}(x) = \int dx^\prime G(x,x^\prime)V(x^\prime)
                 \psi^{(0)}(x^\prime),
\eqn\IIl$$
where
$$
   V(x) = -\xi g^\prime(x)\g_5\g^3,
\eqn\IIm$$
and $G(x,x^\prime)$ is the Green's function satisfying
$$   
  [\ep^2 + {d^2\over{dx^2}} - \xi^2f(x)^2+
  i\xi f^\prime(x)\g^3]_{\a\g}G_{\g\b}(x,x^\prime)
  = - \delta_{\a\b}\delta(x-x^\prime),
\eqn\IIn$$
with the same boundary condition as $\psi^{(0)}(x)$.
Thus, to this order, the spinor solution to the Dirac equation
is given by
$$
 \psi(x)\simeq \e^{-i\s\ep\t}\Bigl\{ [\s\ep\g^0 + 
    i\g^3{d\over{dx}}+\xi f(x)][\psi^{(0)}(x) + \psi^{(1)}(x)]
         - i\xi g(x)\g_5\psi^{(0)}(x) \Bigr\}.
\eqn\IIo$$
\section{Solution to the unperturbed equation}
In the following, any explicit forms of $f(x)$ and $g(x)$ are
not necessary, except for the condition \IIi.
We shall present general results on the wave function
and currents below. The details of the calculation are given in
Appendix.\par
If we expand $\psi^{(0)}(x)$ in terms of the eigenspinors of $\g^3$
as $\psi^{(0)}(x)\sim \phi_\pm(x)u^s_\pm$, with
$$
  \gamma^3 u^s_\pm = \pm i u^s_\pm  \qquad (s=1,2),
\eqn\IIp$$
$\phi_\pm(x)$ must satisfy
$$
  [\ep^2 + {d^2\over{dx^2}} - \xi^2f(x)^2 \mp \xi f^\prime(x)]
  \phi_\pm(x) = 0.
\eqn\IIq$$
Because of \IIi, the asymptotic forms of $\phi_\pm(x)$ should be
$$
 \phi_\pm(x)\rightarrow
 \cases{ \e^{\a x},\  \e^{-\a x} & $(x\rightarrow+\infty)$\cr
         \e^{\b x},\  \e^{-\b x} & $(x\rightarrow-\infty)$\cr}
$$
where $\a=i\sqrt{\ep^2-\xi^2}$ and $\b=i\ep$. Here we consider
the case where the scattering states exist at $x=+\infty$;
$\ep>\xi$. We denote the independent solutions to \IIq\  
as $\phipa_\pm(x)$ and 
$\phima_\pm(x)=\bigl(\phipa_\pm(x)\bigr)^*$ which satisfy
$$\eqalign{
 &\phipa_\pm(x)\rightarrow\e^{\a x},     \cr
 &\phima_\pm(x)\rightarrow\e^{-\a x},   \cr
}\eqn\IIra$$
at $x\rightarrow+\infty$.
Their asymptotic forms at $x\rightarrow-\infty$ are
$$\eqalign{
 &\phipa_\pm(x)\sim \g_\pm(\a,\b)\e^{\b x}+\g_\pm(\a,-\b)\e^{-\b x}, \cr
 &\phima_\pm(x)\sim \g_\pm(-\a,\b)\e^{\b x}+\g_\pm(-\a,-\b)\e^{-\b x},\cr
}\eqn\IIr$$
because of the symmetry of \IIq\  under $\b\mapsto-\b$.
Needless to say, $\g_\pm(\a,\b)^*=\g_\pm(-\a,-\b)$ and
$\g_\pm(\a,-\b)^*=\g_\pm(-\a,\b)$.
With these solutions, the general solution to \IIk\  can be
constructed as
$$
 \psi^{(0)}(x) = \sum_{s=1,2}\sum_\pm
 [(A^s_\pm)^{(-\a)}\phima_\pm(x)+(A^s_\pm)^{(+\a)}\phipa_\pm(x)]
 u^s_\pm.
$$
Multiplied by the operator, $\s\ep\g^0+i\g^3{d\over{dx}}+\xi f(x)$,
it becomes the general solution to the unperturbed (CP-conserving)
Dirac equation,
$$
 \Bigl[\s\ep\g^0+i\g^3{d\over{dx}}-\xi f(x)\Bigr]\psi_0(x)=0.
\eqn\IIs$$
As shown in Appendix, the coefficients, 
$(A^s_-)^{(\pm\a)}, $ are not needed to construct
the general $\psi_0(x)$.
Eventually, the general solution to \IIs\  is given by
$$\eqalign{
 \psi_0(x)=
 &\Bigl[\s\ep\g^0+i\g^3{d\over{dx}}+\xi f(x)\Bigr]\psi^{(0)}(x) \cr
=&\sum_s \Bigl\{
    A_s^{(-)}[\s\ep\phima_+(x)u^s_-+(\xi+\a)\phima_-(x)u^s_+] \cr
 &\qquad\qquad +
    A_s^{(+)}[\s\ep\phipa_+(x)u^s_-+(\xi-\a)\phipa_-(x)u^s_+]
 \Bigr\},  \cr
}\eqn\IIu$$
where
$$
 \psi^{(0)}(x) = \sum_s
 [A_s^{(-)}\phima_+(x) + A_s^{(+)}\phipa_+(x)]u^s_+.
\eqn\IIv$$
This $\psi^{(0)}(x)$ is the starting point of the DWBA.
Any boundary condition can be implemented by choosing 
$A_s^{(-\a)}$ and $A_s^{(+\a)}$ appropriately.
\par
Now we impose a boundary condition to specify the scattering
state. We consider a state in which the incident wave coming
from $x=-\infty$ (symmetric phase) is reflected in part at
the bubble wall, while at $x=+\infty$ only the transmitted wave 
exists.\foot{This boundary condition is referred to as
Type II in ref.[\AJMV].}
This situation is achieved by 
setting $A_s^{(-\a)}=0$ for $\s=+$ and $A_s^{(+\a)}=0$
for $\s=-$, respectively, in \IIu\  and \IIv. 
The Green's function \IIn\  
which matches this boundary condition can be explicitly
constructed from the functions $\phipa_\pm(x)$ and
$\phima_\pm(x)$, as shown in Appendix. Thus
we are ready to compute the currents in the asymptotic
regions.
\section{Wave function and currents in the asymptotic regions}
According to \IIo, the perturbed wave can be obtained once
the unperturbed wave functions and the Green's function
are known. We present only the results, while the derivations
are given in Appendix. The asymptotic forms of the wave
function are
$$\eqalign{
 &\bigl(\psi(x)_{\s=+}\bigr)^{trans}  \cr
=&\e^{-i\ep\t+\a x}\sum_s A_s^{(+)}  \Bigl\{
  (\xi-\a)\Bigl[1+(-)^s{\xi\over\ep}\Bigl({{\xi+\a}\over{2\a}}I_1
     +{1\over2}g(-\infty) \Bigr)\Bigr]u^s_+           \cr
 &\qquad\qquad
     + \ep\Bigl[1+(-)^s{\xi\over\ep}\Bigl({{\xi+\a}\over{2\a}}I_1
        -g(+\infty)+{1\over2}g(-\infty) \Bigr)\Bigr]u^s_- \Bigr\},   \cr
 &\bigl(\psi(x)_{\s=+}\bigr)^{inc}    \cr
=&\e^{-i\ep\t+\b x}\g_+(\a,\b)\sum_s A_s^{(+)} \Bigl\{
   -\b\Bigl[1+(-)^s{\xi\over\ep}\Bigl(
    {\b\over{2\a}}{\g_-(-\a,\b)\over\g_+(\a,\b)}I_2
    +{1\over2}g(-\infty) \Bigr)\Bigr] u^s_+  \cr
 &\qquad\qquad\qquad
  +\ep\bigl[1+(-)^s{\xi\over\ep}\Bigl(
    {\b\over{2\a}}{\g_-(-\a,\b)\over\g_+(\a,\b)}I_2
    - {1\over2}g(-\infty) \Bigr)\Bigr] u^s_-  \Bigr\}, \cr
 &\bigl(\psi(x)_{\s=+}\bigr)^{refl}    \cr
=&\e^{-i\ep\t-\b x}\g_+(\a,-\b)\sum_s A_s^{(+)} \Bigl\{
   \b\Bigl[1+(-)^s{\xi\over\ep}\Bigl(
    -{\b\over{2\a}}{\g_-(-\a,-\b)\over\g_+(\a,-\b)}I_2
    +{1\over2}g(-\infty) \Bigr)\Bigr] u^s_+         \cr
 &\qquad\qquad\qquad
  +\ep\Bigl[1+(-)^s{\xi\over\ep}\Bigl(
    -{\b\over{2\a}}{\g_-(-\a,-\b)\over\g_+(\a,-\b)}I_2
    - {1\over2}g(-\infty) \Bigr)\Bigr] u^s_-  \Bigr\}, \cr
}\eqn\IIw$$
and
$$
  \bigl(\psi(x)_{\s=-}\bigr)^{asym}
 =\bigl(\psi(x)_{\s=+}\bigr)^{asym}
  \Bigr|_{(\a,\b,\ep,I_1,I_2)\rightarrow(-\a,-\b,-\ep,I_1^*,I_2^*)},
\eqn\IIwb$$
where
$$\eqalign{
 I_1 &=\int_{-\infty}^{\infty}dx\, g^\prime(x)\phima_-(x)\phipa_+(x),\cr
 I_2 &=\int_{-\infty}^{\infty}dx\, g^\prime(x)\phipa_-(x)\phipa_+(x).\cr
}\eqn\IIx$$
{}From these, the asymptotic forms of
the vector and axial-vector currents defined by
$$
 j_V^\mu = \bar\psi\g^\mu\psi, \qquad
 j_A^\mu = \bar\psi\g^\mu\g_5\psi,
$$
are (see Appendix)
$$\eqalign{
 &\bigl(j^3_{V,\s}\bigr)^{trans}
= 2\ep\sqrt{\ep^2-\xi^2}\sum_s\abs{A_s^{(\s)}}^2,    \cr
 &\bigl(j^3_{A,\s}\bigr)^{trans}
= 2\ep^2 \sum_s(-)^{s+1}\abs{A_s^{(\s)}}^2.          \cr
 &\bigl(j^3_{V,\s}\bigr)^{inc}_{II}
= 2\ep^2 \abs{\g_+(\a,\b)}^2 \sum_s\abs{A_s^{(\s)}}^2
  [ 1 - \s(-)^{s+1}\delta^{inc} ],  \cr
 &\bigl(j^3_{A,\s}\bigr)^{inc}_{II}
= 2\ep^2 \abs{\g_+(\a,\b)}^2\sum_s(-)^{s+1}\abs{A_s^{(\s)}}^2
  [ 1 - \s(-)^{s+1}\delta^{inc} ],  \cr
 &\bigl(j^3_{V,\s}\bigr)^{refl}_{II}
= -2\ep^2 \abs{\g_+(\a,-\b)}^2 \sum_s\abs{A_s^{(\s)}}^2
  [ 1 + \s(-)^{s+1}\delta^{refl} ],  \cr
 &\bigl(j^3_{A,\s}\bigr)^{refl}_{II}
= 2\ep^2 \abs{\g_+(\a,-\b)}^2\sum_s(-)^{s+1}\abs{A_s^{(\s)}}^2
  [ 1 + \s(-)^{s+1}\delta^{refl} ],  \cr
}\eqn\IIy$$
where the corrections by the CP violation are
$$\eqalign{
 \delta^{inc} &= {\xi\over{2\sqrt{\ep^2-\xi^2}}}  
 \Bigl( {\g_-(-\a,\b)\over\g_+(\a,\b)}I_2 + c.c. \Bigr),    \cr
 \delta^{refl} &= {\xi\over{2\sqrt{\ep^2-\xi^2}}}
 \Bigl( {\g_-(-\a,-\b)\over\g_+(\a,-\b)}I_2 + c.c. \Bigr).    \cr
}\eqn\IIz$$
The chiral currents defined by
$$
  j^\mu_L = {1\over2}(j^\mu_V-j^\mu_A),  \qquad
  j^\mu_R = {1\over2}(j^\mu_V+j^\mu_A),
$$
are
$$\eqalign{
 &\bigl(j^3_{L,\s}\bigr)^{trans}
 = \ep(\sqrt{\ep^2-\xi^2}-\ep)\abs{A^{(\s)}_1}^2
   + \ep(\sqrt{\ep^2-\xi^2}+\ep)\abs{A^{(\s)}_2}^2,  \cr
 &\bigl(j^3_{R,\s}\bigr)^{trans}
 = \ep(\sqrt{\ep^2-\xi^2}+\ep)\abs{A^{(\s)}_1}^2
   + \ep(\sqrt{\ep^2-\xi^2}-\ep)\abs{A^{(\s)}_2}^2,  \cr
 &\bigl(j^3_{L,\s}\bigr)^{inc}
 = 2\ep^2\abs{\g_+(\a,\b)}^2\abs{A^{(\s)}_2}^2(1+\s\delta^{inc}), \cr
 &\bigl(j^3_{R,\s}\bigr)^{inc}
 = 2\ep^2\abs{\g_+(\a,\b)}^2\abs{A^{(\s)}_1}^2(1-\s\delta^{inc}), \cr
 &\bigl(j^3_{L,\s}\bigr)^{refl}
 = -2\ep^2\abs{\g_+(\a,-\b)}^2\abs{A^{(\s)}_1}^2(1+\s\delta^{refl}), \cr
 &\bigl(j^3_{R,\s}\bigr)^{refl}
 = -2\ep^2\abs{\g_+(\a,-\b)}^2\abs{A^{(\s)}_2}^2(1-\s\delta^{refl}). \cr
}\eqn\IIaa$$
\par
With these currents, one can compute the transmission and reflection
coefficients, $T^{(\s)}$ and $R^{(\s)}$, respectively.
{}From the vector currents, we have
$$\eqalign{
 T^{(\s)}
=&{{\bigl(j^3_{V,\s}\bigr)^{trans}}\over{\bigl(j^3_{V,\s}\bigr)^{inc}}}
= T^{(0)}\Bigl[1+{{\sum_s(-)^{s+1}\abs{A^{(\s)}_s}^2}\over
           {\sum_s\abs{A^{(\s)}_s}^2}}\s\delta^{inc}\Bigr], \cr
 R^{(\s)}
=&-{{\bigl(j^3_{V,\s}\bigr)^{refl}}\over{\bigl(j^3_{V,\s}\bigr)^{inc}}}
= R^{(0)}\Bigl[1+{{\sum_s(-)^{s+1}\abs{A^{(\s)}_s}^2}\over
           {\sum_s\abs{A^{(\s)}_s}^2}}
           \s(\delta^{inc}+\delta^{refl})\Bigr], \cr
}\eqn\IIab$$
where
$$\eqalign{
 &T^{(0)}={\a\over\b}{1\over{\abs{\g_+(\a,\b)}^2}},  \cr
 &R^{(0)}=\abs{\g_+(\a,-\b)\over\g_+(\a,\b)}^2,       \cr
}\eqn\IIac$$
are those in the absence of the perturbation.
As shown in Appendix, the unitarity holds:
$$
 T^{(\s)}+R^{(\s)}=T^{(0)}+R^{(0)}=1.
\eqn\IIad$$
The differences in these coefficients between the particle
and the anti-particle would vanish, even in the presence
of CP violation, upon averaged over the initial mixed states
such as the thermal equilibrium.
This is because the difference has a factor, 
${{\sum_s(-)^{s+1}\abs{A^{(\s)}_s}^2}\over{\sum_s\abs{A^{(\s)}_s}^2}}$,
in which $A^{(\s)}_s$ is representation-dependent and can take
any value in a mixed state.
Therefore, it is expected that there would be no net
(vectorlike) particle current through the bubble wall.\par
The transmission and reflection coefficients for the chiral fermion
are defined as
$$\eqalign{
 &T^{(\s)}_{L\rightarrow L(R)} 
 = \bigl(j^3_{L(R),\s}\bigr)^{trans}\bigr|_{A_1^\s=0}
  /\bigl(j^3_{L,\s}\bigr)^{inc},        \cr
 &T^{(\s)}_{R\rightarrow L(R)} 
 = \bigl(j^3_{L(R),\s}\bigr)^{trans}\bigr|_{A_2^\s=0}
  /\bigl(j^3_{R,\s}\bigr)^{inc},         \cr
 &R^{(\s)}_{L(R)\rightarrow R(L)} 
 = -\bigl(j^3_{R(L),\s}\bigr)^{refl}
   /\bigl(j^3_{L(R),\s}\bigr)^{inc}.        \cr
}\eqn\IIae$$
If we denote $R(T)=R^{(+)}(T^{(+)})$ and $\bar R(\bar T)=R^{(-)}(T^{(-)})$,
we have
$$\eqalign{
 &T_{L\rightarrow L}=\bar T_{R\rightarrow R}
={{\sqrt{\ep^2-\xi^2}+\ep}\over{2\ep\abs{\g_+(\a,\b)}^2}}
 (1-\delta^{inc}),     \cr
 &T_{L\rightarrow R}=\bar T_{R\rightarrow L}
={{\sqrt{\ep^2-\xi^2}-\ep}\over{2\ep\abs{\g_+(\a,\b)}^2}}
 (1-\delta^{inc}),     \cr
 &T_{R\rightarrow L}=\bar T_{L\rightarrow R}
={{\sqrt{\ep^2-\xi^2}-\ep}\over{2\ep\abs{\g_+(\a,\b)}^2}}
 (1+\delta^{inc}),     \cr
 &T_{R\rightarrow R}=\bar T_{L\rightarrow L}
={{\sqrt{\ep^2-\xi^2}+\ep}\over{2\ep\abs{\g_+(\a,\b)}^2}}
 (1+\delta^{inc}),     \cr
 &R_{L\rightarrow R}=\bar R_{R\rightarrow L}
=\abs{\g_+(\a,-\b)\over\g_+(\a,\b)}^2
 (1-\delta^{inc}-\delta^{refl}),   \cr
 &R_{R\rightarrow L}=\bar R_{L\rightarrow R}
=\abs{\g_+(\a,-\b)\over\g_+(\a,\b)}^2
 (1+\delta^{inc}+\delta^{refl}).    \cr
}\eqn\IIaf$$
Among these, the following unitarity relations hold:
$$\eqalign{
 &T_{L\rightarrow L}+T_{L\rightarrow R}+R_{L\rightarrow R}=1, \cr
 &T_{R\rightarrow L}+T_{R\rightarrow R}+R_{R\rightarrow L}=1. \cr
}\eqn\IIag$$
The difference between a particle and its anti-particle in
the coefficients is, for example,
$$
 \Delta R
\equiv R_{R\rightarrow L}-\bar R_{R\rightarrow L}
= 2R^{(0)}(\delta^{inc}+\delta^{refl})
=-2T^{(0)}\delta^{inc},
\eqn\IIah$$
where we used $\delta^{inc}+R^{(0)}\delta^{refl}=0$ 
(see Appendix).
This suggests that we would have nonzero chiral-fermion current
through the bubble wall even for the initial state in the 
equilibrium, {\sl if there is a nontrivial CP violation.}
(By `nontrivial', we mean that $g(x)$ does not proportional
to $f(x)$, since in that case the CP-angle can be removed
by a constant bi-unitary transformation of the fermions, or
explicitly, $\delta^{inc}=\delta^{refl}=0$,
as shown  in Appendix.)
\chapter{Application to the Kink-Type Bubble Wall}
The prescription developed in the previous section is applicable
to the situation where the bubble formed in the first-order
phase transition is moving with a constant velocity.
The profile and motion of such a bubble would be determined by
analyzing the detail of dynamics of the phase transition.
Far from the wall, deep in the symmetric/broken phase,
the Higgs field is expected to have the vanishing/nonvanishing 
vacuum expectation value in the uniform vacuum. 
(This corresponds to the condition \IIi.) 
The simplest situation may be that in which
the profile has the shape of the kink. This case,
without CP violation, was studied in ref.[\AJMV].
Now we apply our procedure to that case, and evaluate
the transmission and reflection coefficients.
\par
Following [\AJMV], we suppose that the bubble-wall profile
is given by
$$
   f(x)={{1+\tanh x}\over2}.
\eqn\IIIa$$
Then the solutions to the Klein-Gordon-type equation, \IIq,
are expressed in terms of the hypergeometric functions:
$$\eqalign{
  \phipa_\pm(y) 
&=y^{-\a/2}(1-y)^{\b/2} 
  F({{-\a+\b\mp\xi}\over2}+1, {{-\a+\b\pm\xi}\over2}, -\a+1;y), \cr
  \phima_\pm(y) 
&=y^{\a/2}(1-y)^{\b/2} 
  F({{\a+\b\mp\xi}\over2}+1, {{\a+\b\pm\xi}\over2}, \a+1;y)
 =\bigl(\phipa_\pm(y)\bigr)^*, \cr
}\eqn\IIIb$$
where we have changed the variable as $y=(1-\tanh x)/2$.
\foot{The parameters $\a$, $\b$ and $\xi$ in ref.[\AJMV] are
halves of ours.}
These have the desired asymptotic behaviors as \IIra\  and \IIr. 
Now $\g_\pm(\a,\b)$ are explicitly given by
$$
 \g_\pm(\a,\b)
={{\G(-\a+1)\G(-\b)}\over
  {\G({{-\a-\b\pm\xi}\over2})\G({{-\a-\b\mp\xi}\over2}+1)} }.
\eqn\IIIe$$
The transmission and reflection coefficients without the CP 
violation are
$$\eqalign{
 &T^{(0)}
={{\sin(\pi\a)\sin(\pi\b)}\over
  {\sin[{\pi\over2}(\a+\b+\xi)]\sin[{\pi\over2}(\a+\b-\xi)]}},\cr
 &R^{(0)}
={{\sin[{\pi\over2}(\a-\b+\xi)]\sin[{\pi\over2}(\a-\b-\xi)]}\over
  {\sin[{\pi\over2}(\a+\b+\xi)]\sin[{\pi\over2}(\a+\b-\xi)]}}.\cr
}\eqn\IIIf$$
\par
The effects of the CP violation can be evaluated, once
the functional form of $g(x)$ is given. One of the quantities
we concern may be the difference in the reflection/transmission
coefficients between the particle and anti-particle.
As mentioned above, the vectorlike one would vanish
after averaging over the initial mixed state.
However, the chiral one, which is given by \IIah, would survive 
even for the initial state in the equilibrium.
\par
We have executed numerical calculations for various $g(x)$.
Here we present the results for two typical cases in
which $g^\prime(x)$ has different behaviors around the wall surface,
{\it i.e.,} $g(x)=\Delta\theta\cdot f(x)^2$ and
$g(x)=\Delta\theta{d\over{dx}}f(x)$, where $\Delta\theta$ 
characterizes the magnitude of CP violation.
We carried out the numerical calculation of
$\Delta R$ at several energies. The $a$-dependence of 
$\Delta R/\Delta\theta$ for $g(x)=\Delta\theta\cdot f(x)^2$
is summarized in Table 1.
The $E^*$-dependence of $\Delta R/\Delta\theta$ is plotted 
in Figs.~1 and 2 for the two choices of $g(x)$ for several $a$.
{}From these, we find the following:
\item{(1)} The sign of $\Delta R$ varies not only with $a$ 
but also with the functional form of $g(x)$.
\item{(2)} The absolute value of $\Delta R$ rapidly decreases
as the energy of the incident particle increases. 
This feature may be independent of the choice of $g(x)$, since the 
$g(x)$-independent factor of $\Delta R$ is very small in the
region far from $E^*\sim m_0$. 
\item{(3)} The degree of the decrease of $\abs{\Delta R}$ seems to 
be weakened as $a$ increases, {\it i.e.} the bubble wall becomes thinner.
\chapter{Conclusions}
We have developed a general prescription to solve the Dirac
equation in a bubble-wall background with CP violation, 
to the first order of CP violation. The transmission and reflection
coefficients are obtained to this order, in terms of
the functions of the energy, the mass in the broken
phase and the wall thickness, which are extracted from
the continuation of the solution to the second-order
Klein-Gordon-like equation. 
It is found, irrespectively of the detailed form of
the bubble-wall profile, that for the vectorlike current, 
the effect of CP violation will be canceled, when
averaged over an initial mixed state such as thermal 
equilibrium, and that the net chiral current through the
bubble wall would remain.
Since the relation between the chemical potentials 
of the left- and right-handed quarks in the symmetric 
phase is different from that in the broken phase
\refmark\HT, this would leave net baryon number
after the phase transition.
\par
We have applied our prescription to the simple example in which
the profile of the bubble wall is of the kink type.
This situation may be realized when the phase
transition proceeds slowly with degenerate minima
of the free energy.
Some numerical results are obtained in the case where the 
CP-violating angle has nontrivial spatial dependence.
$\Delta R$, which measures the difference in the reflection
probabilities between the chiral fermion and anti-fermion, is 
evaluated for various energies and wall thickness, assuming
some typical forms of $g(x)$. The results suggest that
irrespectively of the choice of $g(x)$, $\abs{\Delta R}$ rapidly
decreases as the energy of incident fermions increases.
Further the degree of decreasing seems to be larger
as the wall becomes thicker.\par
In order to determine whether and how much the net
baryon number is left after the electroweak phase transition,
we must know the detailed dynamics of the phase transition
and particle distribution within both phases divided
by the bubble wall. 
The baryon number left in the broken phase may be
calculated by taking the average of the net fermion current
through the bubble wall from the symmetric to the
broken phase, and vice versa. Since the distribution function
of the chiral fermions in the broken phase is different
from that in the symmetric phase, as shown from the
equilibrium analysis in [\HT], a quantity
such as $\Delta R$ will play an important role in
estimating the baryon asymmetry of the universe.
As far as the cases we have studied are concerned,
$\Delta R$ shows no drastic enhancement.
On the other hand, even the sign of $\Delta R$ varies with 
the inverse wall-thickness $a$, as well as with the functional 
form of $g(x)$. 
Thus whether the baryon or anti-baryon is left after 
the electroweak phase transition would be a subtle problem. 
We would like to remark that $\abs{\Delta\theta}$ does not need
to be extremely small in so far as $\abs{g(+\infty)}$ is small
enough to satisfy the experimental bound.
In fact, in our second example ($g(x)=\Delta\theta f^\prime(x)$),
$g(+\infty)$ vanishes irrespective of the value of $\Delta\theta$.
This fact might allow us to build a model which produces
sufficient baryon number, by taking a suitable form of $g(x)$.
\par
Our results will be applied to the scenario of the electroweak
baryogenesis, which assumes the first-order phase transition.
Although much should be revealed about the mechanism of CP
violation and the dynamics of the
phase transition, such as the order of it and the motion of
the bubble wall when it is of first order, the baryon asymmetry
of the universe might be generated in the electroweak phase
transition era if there is sufficient baryon number current
through the bubble wall.
\appendix
In this appendix, we shall derive some equations used 
in the text.\par\noindent
\line{\it Unperturbed wave function\hfill}
The second-order unperturbed equation \IIq\  is written as
$$
 D_\mp D_\pm\phi_\pm(x)=\ep^2\phi_\pm(x),
\eqn\Aa$$
where $\phi(x)$ is either $\phipa(x)$ or $\phima(x)$ and
the operator $D_\pm$ is defined by
$$
  D_\pm \equiv \mp{d\over{dx}}+\xi f(x).
\eqn\Ab$$
Multiplying \Aa\  by $D_\pm$, we have
$$
 \bigl(D_\pm D_\mp\bigr)\bigl(D_\pm\phipa_\pm(x)\bigr)
=\ep^2 \bigl(D_\pm\phipa_\pm(x)\bigr).
$$
Since the independent solutions to this equation are
$\phipa_\mp(x)$ and $\phima_\mp(x)$, 
$D_\pm\phipa_\pm(x)$ should be expressed as
$$
 D_\pm\phipa_\pm(x)
= c^{(+\a)}_\pm\phipa_\mp(x)+c^{(-\a)}_\pm\phima_\mp(x).
$$
{}From the behavior of the left-hand side at $x\sim+\infty$;
$$
 D_\pm\phipa_\pm(x)\sim (\xi\mp\a)\e^{\a x},
$$
it follows that $c^{(+\a)}_\pm=\xi\mp\a$ and $c^{(-\a)}_\pm=0$.
That is,
$$\eqalign{
[\mp{d\over{dx}}+\xi f(x)]\phima_\pm(x) &=(\xi\pm \a)\phima_\mp(x), \cr
[\mp{d\over{dx}}+\xi f(x)]\phipa_\pm(x) &=(\xi\mp \a)\phipa_\mp(x). \cr
}\eqn\Ac$$
In the Dirac representation, the $\g$-matrices are represented as
$$
 \g^3 = \pmatrix{ 0 & \s_3 \cr -\s_3 & 0 \cr},\qquad
 \g^0=\pmatrix{ 1 & 0 \cr 0 & -1 \cr},
$$
and the eigenspinors of $\g^3$ satisfying
$$
   \g^3 u^s_\pm = \pm i u^s_\pm,  \qquad (s=1,2)
\eqn\Ad$$
are expressed as
$$
 u^1_\pm = {1\over\sqrt{2}}\pmatrix{1\cr 0\cr \pm i\cr 0\cr},\qquad
 u^2_\pm = {1\over\sqrt{2}}\pmatrix{0\cr 1\cr 0\cr \mp i\cr}.
\eqn\Ae$$
Then $\g^0 u^s_\pm = u^s_\mp$.
The general solution to the unperturbed Dirac equation \IIs\  
is obtained by multiplying the most general
$\psi^{(0)}(x)$ by the operator 
$\s\ep\g^0+i\g^3{d\over{dx}}+\xi f(x)$:
$$\eqalign{
 \psi_0(x)
&\equiv [\s\ep\g^0+i\g^3{d\over{dx}}+\xi f(x)]\psi^{(0)}(x) \cr
&=\sum_s \Bigl\{
  (A^s_+)^{(-\a)}[\s\ep\phima_+(x)u^s_-+(\xi+\a)\phima_-(x)u^s_+] \cr
&\qquad\quad
 +(A^s_-)^{(-\a)}[\s\ep\phima_-(x)u^s_++(\xi-\a)\phima_+(x)u^s_-] \cr 
&\qquad\quad 
 +(A^s_+)^{(+\a)}[\s\ep\phipa_+(x)u^s_-+(\xi-\a)\phipa_-(x)u^s_+] \cr 
&\qquad\quad 
 +(A^s_-)^{(+\a)}[\s\ep\phipa_-(x)u^s_++(\xi+\a)\phipa_+(x)u^s_-] 
 \Bigr\}.  \cr 
}\eqn\Af$$
Here $\s\ep\phima_+(x)u^s_- +(\xi+\a)\phima_-(x)u^s_+$ and
$\s\ep\phima_-(x)u^s_+ +(\xi-\a)\phima_+(x)u^s_-$ are not
independent with each other, since the determinant of the
coefficients of their components vanishes:
$\left|\,\matrix{\s\ep&\xi+\a\cr \xi-\a&\s\ep\cr}\right|=0.$ 
Similarly the functions in the
last two lines of \Af\  are not independent. Hence the most
general solution is given by putting 
$(A^s_-)^{(-\a)}=(A^s_-)^{(+\a)}=0$; the most general
$\psi^{(0)}(x)$ is
$$
 \psi^{(0)}(x) = \sum_s
 [A_s^{(-\a)}\phima_+(x) + A_s^{(+\a)}\phipa_+(x)]u^s_+.
\eqn\Ag$$
The wave function satisfying the boundary condition which we
adopted in the text is obtained by putting $A_s^{(-\s\a)}=0$.
\par\noindent
\line{\it Green's function\hfill}
Here we construct the Green's function satisfying
$$   
  [\ep^2 + {d^2\over{dx^2}} - \xi^2f(x)^2+
  i\xi f^\prime(x)\g^3]_{\a\g}G_{\g\b}(x,x^\prime)
  =\Delta_{\a\g}G_{\g\b}(x,x^\prime)
  = - \delta_{\a\b}\delta(x-x^\prime),
\eqn\Ah$$
with the boundary condition discussed in the text.
Put
$$
 \Delta_\pm 
={d^2\over{dx^2}}\mp\xi f^\prime(x) - \xi^2f(x)^2+\ep^2, 
\eqn\Ai$$
and introduce a unitary matrix defined by
$$
 U = (u^1_+\  u^1_-\  u^2_+\  u^2_-).
$$
Then, since $\Delta$ is diagonalized as
$$
 \Delta=U\pmatrix{\Delta_+& & & \cr  &\Delta_-& & \cr
             & &\Delta_+& \cr & & &\Delta_-\cr}U^{-1},
$$
the Green's function is given by
$$
  G(x,x^\prime)
 =U  \pmatrix{G_+(x,x^\prime)&  &  & \cr
               &G_-(x,x^\prime)&  &  \cr
               &  &G_+(x,x^\prime)&  \cr
               &  &  &G_-(x,x^\prime)\cr} U^{-1},
$$
where $G_\pm(x,x^\prime)$ satisfies
$$
  \Delta_\pm G_\pm(x,x^\prime) = -\delta(x-x^\prime),
\eqn\Aj$$
with the same boundary condition as $G(x,x^\prime)$.
One can construct the Green's function $G_\pm(x,x^\prime)$
following the standard method\refmark\CH. 
If $u_\pm(x)$ ($v_\pm(x)$) are solutions to
$\Delta_\pm u_\pm(x)=0$($\Delta_\pm v_\pm(x)=0$) and
satisfy the boundary condition imposed on $G(x,x^\prime)$
at $x\rightarrow-\infty$($x\rightarrow+\infty$), then
the Green's function is expressed as
$$
 G_\pm(x,x^\prime)
=\cases{ -{1\over{W(u_\pm,v_\pm)}}u_\pm(x)v_\pm(x^\prime),
         & $(x<x^\prime)$ \cr
         -{1\over{W(u_\pm,v_\pm)}}u_\pm(x^\prime)v_\pm(x),
         & $(x^\prime<x)$ \cr}
\eqn\Ak$$
where $W(u,v)$ is the Wronskian defined by
$$
 W(u_\pm,v_\pm)
=u_\pm(x){d\over{dx}}v_\pm(x)-v_\pm(x){d\over{dx}}u_\pm(x)
={\rm const.}
\eqn\Al$$
For the boundary condition in question, there exists only
right-moving wave at $x\sim+\infty$ so that we should choose,
for each $\s$,
$$
  v_\pm(x) = \phipsa_\pm(x),
$$
while we take the most general choice for $u_\pm(x)$;
$$
 u_\pm(x)=\phimsa_\pm(x)+c^{(\s)}_\pm\phipsa_\pm(x).
$$
where $c^{(\s)}_\pm$ is some constant.
(The overall normalization is irrelevant, since the product
of $u$ and $v$ is divided by the Wronskian in \Ak.)
By use of the asymptotic forms of $\phi_\pm(x)$ 
at $x\sim+\infty$,
$$
 W(u_\pm,v_\pm)=2\s\a.
$$
Thus the Green's function $G_\pm(x,x^\prime)$ is
$$
 G_\pm^{(\s)}(x,x^\prime)
=\cases{ -{\s\over{2\a}}
         [\phimsa_\pm(x)+c^{(\s)}_\pm\phipsa_\pm(x)]
         \phipsa_\pm(x^\prime),  & $(x<x^\prime)$ \cr
         -{\s\over{2\a}}
         [\phimsa_\pm(x^\prime)+c^{(\s)}_\pm\phipsa_\pm(x^\prime)]
         \phipsa_\pm(x),  & $(x^\prime<x)$ \cr}.
\eqn\Am$$
\par\noindent
\line{\it DWBA to the first order\hfill}
As explained in Section 2.2, the function $\psi_\ep(x)$
in the Ansatz \IIg\  is given by
$$
   \psi_\ep(x) = \psi^{(0)}(x) + \psi^{(1)}(x),
\eqn\An$$
where $\psi^{(0)}(x)$ is a solution to the unperturbed equation
with the boundary condition above, that is,
$$
 \psi^{(0)}(x)=\sum_s A_s^{(\s)}\phipsa_+(x)u^s_+,
\eqn\Ao$$
and $\psi^{(1)}(x)$ is
$$
 \psi^{(1)}(x) = \int dx^\prime G(x,x^\prime)V(x^\prime)
                 \psi^{(0)}(x^\prime),
\eqn\Ap$$
Noting that in the Dirac representation, 
$\gamma_5 = \pmatrix{0&1\cr 1&0\cr}$ so that
$$
  \gamma_5\gamma^3 u^s_\pm = (-)^s u^s_\mp,
$$
and $V(x)$ is 
$$
   V(x) = -\xi g^\prime(x)\g_5\g^3,
$$
the integrand of \Ap\  is written as
$$
  G(x,x^\prime) V(x^\prime) \psi^{(0)}(x^\prime)
= -\xi G^{(\s)}_-(x,x^\prime) g^\prime(x^\prime)
   \sum_s (-)^sA_s^{(\s)}\phipsa_+(x^\prime) u^s_-.
$$
Therefore $\psi^{(1)}(x)$ is 
$$\eqalign{
 \psi^{(1)}(x)
=& {{\s\xi}\over{2\a}}\sum_s (-)^sA_s^{(\s)}u^s_-  \cr
 &\times \Bigl\{
  [\phimsa_-(x)+c^{(\s)}_-\phipsa_-(x)]
  \int_x^\infty dx^\prime\, g^\prime(x^\prime)
  \phipsa_-(x^\prime)\phipsa_+(x^\prime)       \cr
 &+\phipsa_-(x)\int^x_{-\infty}dx^\prime\, g^\prime(x^\prime)
  [\phimsa_-(x^\prime)+c^{(\s)}_-\phipsa_-(x^\prime)]
   \phipsa_+(x^\prime) \Bigr\}   \cr
}$$
and its asymptotic forms are
$$\eqalign{
  \bigl(\psi^{(1)}(x)\bigr)^{+\infty}
=&{\xi\over{2\a}}\e^{\a x}(I_1+c^{(+)}_-I_2)
  \sum_s (-)^s A_s^{(+)}u^s_-,   \cr
  \bigl(\psi^{(1)}(x)\bigr)^{-\infty}
=&{\xi\over{2\a}}I_2 \sum_s (-)^s A_s^{(+)}u^s_-  \cr
 &\times \bigl[
  \bigr(\g_-(-\a,\b)+c^{(+)}_-\g_-(\a,\b)\bigr)\e^{\b x}
 +\bigr(\g_-(-\a,-\b)+c^{(+)}_-\g_-(\a,-\b)\bigr)\e^{-\b x}
  \bigr],   \cr
}$$
for the positive-energy wave, and
$$\eqalign{
  \bigl(\psi^{(1)}(x)\bigr)^{+\infty}
=&-{\xi\over{2\a}}\e^{-\a x}(I_1^*+c^{(-)}_-I_2^*)
  \sum_s (-)^s A_s^{(-)}u^s_-,   \cr
  \bigl(\psi^{(1)}(x)\bigr)^{-\infty}
=&-{\xi\over{2\a}}I_2^* \sum_s (-)^s A_s^{(-)}u^s_-  \cr
 &\times \bigl[
  \bigr(\g_-(\a,\b)+c^{(-)}_-\g_-(-\a,\b)\bigr)\e^{\b x}
 +\bigr(\g_-(\a,-\b)+c^{(-)}_-\g_-(-\a,-\b)\bigr)\e^{-\b x}
  \bigr],   \cr
}$$
for the negative-energy wave, where
$$\eqalign{
 I_1 &= \int_{-\infty}^{\infty}dx\, g^\prime(x)
        \phima_-(x)\phipa_+(x) ,   \cr
 I_2 &= \int_{-\infty}^{\infty}dx\, g^\prime(x)
        \phipa_-(x)\phipa_+(x).    \cr
}\eqn\Aq$$
With these asymptotic forms and those of $\psi^{(0)}(x)$,
$$\eqalign{
 &\bigl(\psi^{(0)}(x)\bigr)^{+\infty}
 =\e^{\s\a x}\sum_s A_s^{(\s)} u^s_+,   \cr
 &\bigl(\psi^{(0)}(x)\bigr)^{-\infty}
 =\sum_s A_s^{(\s)} u^s_+
  [\g_+(\s\a,\b)\e^{\b x}+\g_+(\s\a,-\b)\e^{-\b x}],  \cr
}$$
we can compute the asymptotic forms of the perturbed 
wave function using \IIo;
$$
 \psi(x)= \e^{-i\s\ep\t}\Bigl\{ [\s\ep\g^0 + 
    i\g^3{d\over{dx}}+\xi f(x)][\psi^{(0)}(x) + \psi^{(1)}(x)]
         - i\xi g(x)\g_5\psi^{(0)}(x) \Bigr\}.
\eqn\Ar$$
For the positive-energy wave, we have
$$\eqalign{
  \bigl(\psi(x)_{\s=+}\bigr)^{trans}
=&\e^{-i\ep\t+\a x}\sum_s A_s^{(+)}
  \Bigl[1+(-)^s{\xi\over\ep}{{\xi+\a}\over{2\a}}c^{(+)}_-I_2\Bigr] \cr
 &\times\Bigl\{
  (\xi-\a)\Bigl[1+(-)^s{\xi\over\ep}{{\xi+\a}\over{2\a}}I_1
               \Bigr]u^s_+           \cr
 &\qquad + \ep\Bigl[1+(-)^s{\xi\over\ep}\Bigl({{\xi+\a}\over{2\a}}I_1
        -g(+\infty) \Bigr)\Bigr]u^s_- \Bigr\},   \cr
  \bigl(\psi(x)_{\s=+}\bigr)^{inc}
=&\e^{-i\ep\t+\b x}\g_+(\a,\b)\sum_s A_s^{(+)} 
  \Bigl[1+(-)^s{\xi\over\ep}{\b\over{2\a}}
        {\g_-(\a,\b)\over\g_+(\a,\b)}c^{(+)}_-I_2\Bigr] \cr
 &\times\Bigl\{
   -\b\Bigl[1+(-)^s{\xi\over\ep}
    {\b\over{2\a}}{\g_-(-\a,\b)\over\g_+(\a,\b)}I_2
     \Bigr] u^s_+  \cr
 &\qquad
  +\ep\bigl[1+(-)^s{\xi\over\ep}\Bigl(
    {\b\over{2\a}}{\g_-(-\a,\b)\over\g_+(\a,\b)}I_2
    - g(-\infty) \Bigr)\Bigr] u^s_-  \Bigr\}, \cr
  \bigl(\psi(x)_{\s=+}\bigr)^{refl}
=&\e^{-i\ep\t-\b x}\g_+(\a,-\b)\sum_s A_s^{(+)} 
  \Bigl[1-(-)^s{\xi\over\ep}{\b\over{2\a}}
        {\g_-(\a,-\b)\over\g_+(\a,-\b)}c^{(+)}_-I_2\Bigr] \cr
 &\times\Bigl\{
   \b\Bigl[1-(-)^s{\xi\over\ep}
    {\b\over{2\a}}{\g_-(-\a,-\b)\over\g_+(\a,-\b)}I_2
     \Bigr] u^s_+         \cr
 &\qquad
  +\ep\Bigl[1+(-)^s{\xi\over\ep}\Bigl(
    -{\b\over{2\a}}{\g_-(-\a,-\b)\over\g_+(\a,-\b)}I_2
    - g(-\infty) \Bigr)\Bigr] u^s_-  \Bigr\}. \cr
}\eqn\As$$
The negative-energy wave is related to the positive one by
$$
  \bigl(\psi(x)_{\s=-}\bigr)^{asym}
= \bigl(\psi(x)_{\s=+}\bigr)^{asym}
  \Bigr|_{(\a,\b,\ep,I_1,I_2)\rightarrow(-\a,-\b,-\ep,I_1^*,I_2^*)}.
$$
When one of the relations \Ac, 
$$
 \bigl[{d\over{dx}}+\xi f(x)\bigr]\phipa_-(x)
=(\xi+\a)\phipa_+(x),
$$
is evaluated at $x\sim-\infty$, it produces
$$
 \b\g_-(\a,\b)\e^{\b x}-\b\g_-(\a,-\b)\e^{-\b x}
=(\xi+\a)[\g_+(\a,\b)\e^{\b x}+\g_+(\a,-\b)\e^{-\b x}],
$$
or
$$\eqalign{
 &\b\g_-(\a,\b)=(\xi+\a)\g_+(\a,\b),   \cr
 &\b\g_-(\a,-\b)=-(\xi+\a)\g_+(\a,-\b).   \cr
}\eqn\At$$
These relations lead to
$$
 {{\xi+\a}\over{2\a}}
={\b\over{2\a}}{\g_-(\a,\b)\over\g_+(\a,\b)}
=-{\b\over{2\a}}{\g_-(\a,-\b)\over\g_+(\a,-\b)}.
$$
Hence all the factors including $c^{(+)}_-I_2$ in \As\  are
the same, so that they are absorbed in the arbitrary coefficients
$A_s^{(\pm)}$ by redefining them as
$$\eqalign{
&A_s^{(+)}
 \Bigl[1+(-)^s{\xi\over\ep}{{\xi+\a}\over{2\a}}c^{(+)}_-I_2\Bigr]
 \Bigl[1-(-)^s{\xi\over\ep}{1\over2}g(-\infty)\Bigr]
 \rightarrow A_s^{(+)},  \cr
&A_s^{(-)}
 \Bigl[1+(-)^s{\xi\over\ep}{{\xi-\a}\over{2\a}}c^{(-)}_-I_2^*\Bigr]
 \Bigl[1+(-)^s{\xi\over\ep}{1\over2}g(-\infty)\Bigr]
 \rightarrow A_s^{(-)},  \cr
}$$
where the last factor of the left-hand side is put for
later convenience.
This gives the result \IIw.
\par\noindent
\line{\it Currents and unitarity\hfill}
Now it is straightforward to calculate the asymptotic forms
of the various currents, with the wave function \IIw.
The straightforward calculation yields the same result as \IIy\  
except for $\bigl(j^3_{V,\s}\bigr)^{trans}$ and
$\bigl(j^3_{A,\s}\bigr)^{trans}$, which are
$$\eqalign{
 &\bigl(j^3_{V,\s}\bigr)^{trans}
 =-2i\ep\a\sum_s\abs{A_s^{(\s)}}^2\Bigl\{ 1 + (-)^s\s{\xi\over\ep}
  \Bigl[-g(+\infty)+g(-\infty)+
    \Bigl({{\xi+\a}\over{2\a}}I_1 + c.c.\Bigr)\Bigr]\Bigr\}, \cr
 &\bigl(j^3_{A,\s}\bigr)^{trans}
 =2\ep^2\sum_s (-)^s\abs{A_s^{(\s)}}^2\Bigl\{ 1 + (-)^s\s{\xi\over\ep}
  \Bigl[-g(+\infty)+g(-\infty)+
    \Bigl({{\xi+\a}\over{2\a}}I_1 + c.c.\Bigr)\Bigr]\Bigr\}. \cr
}$$
Here one can show that
$$
 {{\xi+\a}\over{2\a}}I_1 + c.c.=  g(+\infty)-g(-\infty),
\eqn\Au$$
so that the result in \IIy\  is obtained. Let us prove this 
equation. Define $\tilde g(x)=g(x)-g(-\infty)$, then
$g^\prime(x)=\tilde g^\prime(x)$ and $\tilde g(-\infty)=0$.
Integrating by parts, $I_1$ becomes
$$\eqalign{
 I_1
=&\int_{-\infty}^{\infty}dx\, \tilde g^\prime(x)
        \phima_-(x)\phipa_+(x)  \cr
=&\tilde g(x)\phima_-(x)\phipa_+(x) \bigr|_{-\infty}^\infty  \cr
 &\qquad
  -\int_{-\infty}^{\infty}dx\, \tilde g(x)
   \bigl[ {\phima_-}^\prime(x)\phipa_+(x)
         +\phima_-(x){\phipa_+}^\prime(x) \bigr].  \cr
}$$
{}From \Ac, the first derivatives of $\phi_\pm(x)$ can be written as
$$\eqalign{
 &{\phima_-}^\prime(x)
 =-\xi f(x)\phima_-(x)+(\xi-\a)\phima_+(x),  \cr
 &{\phipa_+}^\prime(x)
 = \xi f(x)\phipa_+(x)-(\xi-\a)\phipa_-(x).  \cr
}$$
Hence,
$$
 I_1=\tilde g(+\infty) 
    -(\xi-\a)\int_{-\infty}^{\infty}dx\, \tilde g(x)
     \bigl[\abs{\phipa_+(x)}^2-\abs{\phipa_-(x)}^2 \bigr],
$$
which leads to
$$\eqalign{
 &{{\xi+\a}\over{2\a}}I_1 + c.c.   \cr
=&\Bigl({{\xi+\a}\over{2\a}}+c.c.\Bigr)\tilde g(+\infty)
 -\Bigl({{\xi^2-\a^2}\over{2\a}}+c.c.\Bigr)
  \int_{-\infty}^{\infty}dx\, \tilde g(x)
     \bigl[\abs{\phipa_+(x)}^2-\abs{\phipa_-(x)}^2 \bigr] \cr
=&\tilde g(+\infty).  \cr
}$$
\par
Next we shall examine the unitarity.
Without the perturbation, the unitarity relation,
$$
  T^{(0)}+R^{(0)}
 ={\a\over\b}{1\over{\abs{\g_+(\a,\b)}^2}}
  +\abs{\g_+(\a,-\b)\over\g_+(\a,\b)}^2
 =1,
\eqn\Av$$
immediately follows from the current conservation which is
derived from the Klein-Gordon-type equation, \IIq,
or in other words, from the fact that the Wronskian 
$W(\phima_+,\phipa_+)$ is independent of $x$.
 Evaluated at $x\sim+\infty$ and at $x\sim-\infty$, 
the Wronskian is
$$
 W(\phima_+,\phipa_+)=2\a
=2\b\bigl(\abs{\g_+(\a,\b)}^2-\abs{\g_+(\a,-\b)}^2\bigr),
$$  
which is nothing but \Av.
In the presence of the CP-violation, the sum of the transmission
and reflection coefficients is, from \IIab,
$$
 T^{(\s)}+ R^{(\s)}
=1+
 {{\sum_s(-)^{s+1}\abs{A^{(\s)}_s}^2}\over{\sum_s\abs{A^{(\s)}_s}^2}}
 \s(\delta^{inc}+R^{(0)}\delta^{refl}).
$$
Now we can show that
$$
  \delta^{inc}+R^{(0)}\delta^{refl}=0,
\eqn\Aw$$
so that the unitarity of the perturbed wave holds. 
This relations is verified as follows. From the definitions,
$$
 \delta^{inc}+R^{(0)}\delta^{refl}
={\xi\over\ep}{\b\over{2\a}}\Bigl[
 \Bigl({\g_-(-\a,\b)\over\g_+(\a,\b)}
      +R^{(0)}{\g_-(-\a,-\b)\over\g_+(\a,-\b)}\Bigr)
  + c.c.\Bigr].
$$
Here the term in the parentheses is written as
$$
 {\g_-(-\a,\b)\over\g_+(\a,\b)}+R^{(0)}{\g_-(-\a,-\b)\over\g_+(\a,-\b)}
={{\g_+(-\a,-\b)\g_-(-\a,\b)+\g_+(-\a,\b)\g_-(-\a,-\b)}
  \over{\abs{\g_+(\a,\b)}^2}}=0,
$$
where we have used \At\  to rewrite $\g_-$ in terms of
$\g_+$ in the denominator.\par
The transmission and reflection coefficients 
of the chiral fermions are expressed as
$$\eqalign{
 &T_{L\rightarrow L}+T_{L\rightarrow R}
 =T^{(0)}(1-\delta^{inc}),  \cr
 &T_{R\rightarrow L}+T_{R\rightarrow R}
 =T^{(0)}(1+\delta^{inc}),  \cr
 &R_{L\rightarrow R}=R^{(0)}(1-\delta^{inc}-\delta^{refl}),\cr
 &R_{R\rightarrow L}=R^{(0)}(1+\delta^{inc}+\delta^{refl}),\cr
}$$
from which it follows that
$$\eqalign{
 &T_{L\rightarrow L}+T_{L\rightarrow R}+R_{L\rightarrow R}
=1-(\delta^{inc}+R^{(0)}\delta^{refl})
=1, \cr
 &T_{R\rightarrow L}+T_{R\rightarrow R}+R_{R\rightarrow L}
=1+(\delta^{inc}+R^{(0)}\delta^{refl})
=1, \cr
}$$
because of \Aw.\par
When $g(x)$ is a linear function of $f(x)$, it is shown that
$\delta^{inc}=\delta^{refl}=0$. This is verified as follows.
By use of the differential equation \Aa, the integral in which 
$g^\prime(x)$ is replaced with $f^\prime(x)$ becomes
$$\eqalign{
\tilde I_2 \equiv
 &\int_{-\infty}^{\infty}dx\, f^\prime(x)\phipa_-(x)\phipa_+(x) \cr
=&{1\over{2\xi}}\int_{-\infty}^{\infty}dx\, 
     \Bigl[ \phipa_-(x){d^2\over{dx^2}}\phipa_+(x)
           -\phipa_+(x){d^2\over{dx^2}}\phipa_-(x) \Bigr]   \cr
=&{1\over{2\xi}}
     \Bigl[ \phipa_-(x){d\over{dx}}\phipa_+(x)
      -\phipa_+(x){d\over{dx}}\phipa_-(x) \Bigr]_{-\infty}^{\infty} \cr
=&-{\b\over\xi} [\g_+(\a,\b)\g_-(\a,-\b)-\g_+(\a,-\b)\g_-(\a,\b)].  \cr
}$$
Then, with the help of \At,
$$
 {\g_-(-\a,\b)\over\g_+(\a,\b)}\tilde I_2
=-{{2\b}\over\xi}\abs{\g_-(\a,-\b)}^2,
$$
which is purely imaginary, so that $\delta^{inc}=0$ from the definition
in \IIz, and $\delta^{refl}=0$ because of \Aw.
\refout
\vfill\eject
%
%   Here comes the captions
%
%
\frontpagetrue
\centerline{\fourteenbf Table Caption}
\item{\bf Table 1.} The values of $\Delta R/\Delta\theta$
for various energy of the incident particle and
thickness of the bubble wall. We have chosen
$g(x) = \Delta\theta f(x)^2$.
The numerical values of $E^*$ and $a$ are given in the unit of
$m_0$, the height of the wall.
\bigskip
\centerline{\fourteenbf Figure Captions}
\item{\bf Figure 1.} $\Delta R/\Delta\theta$ as a function of
$E^*$ for various $a$, in the case where $g(x)=\Delta\theta f(x)^2$.
The numerical values of $E^*$ and $a$ are given in the unit of
$m_0$, the height of the wall.
\item{\bf Figure 2.} $\Delta R/\Delta\theta$ as a function of
$E^*$ for various $a$, in the case where 
$g(x)=\Delta\theta {d\over{dx}}f(x)$.
The numerical values of $E^*$ and $a$ are given in the unit of
$m_0$, the height of the wall.
%
%
%   Here comes the table.
%
\bigskip
\begintable
$a$ | $E^*=1.001$ |    $1.5$        |    $2.0$         |    $5.0$    \crthick
0.1 |$0.152$      |\scien{2.32}{-18}|\scien{-2.42}{-26}|\scien{-2.42}{-70}\cr
0.2 |$0.387$      |\scien{9.12}{-10}|\scien{4.96}{-14} |\scien{4.69}{-36} \cr
0.4 |$0.277$      |\scien{5.98}{-6} |\scien{1.07}{-8}  |\scien{-6.23}{-20}\cr
0.6 |$-1.14$      |\scien{8.71}{-6} |\scien{-2.06}{-7} |\scien{-1.15}{-13}\cr
0.8 |$-0.640$     |\scien{-1.00}{-3}|\scien{-6.50}{-5} |\scien{-1.37}{-10}\cr
1.0 |$-0.474$     |\scien{-3.12}{-3}|\scien{-3.19}{-4} |\scien{-6.94}{-9} \cr
2.0 |$-0.244$     |\scien{-1.61}{-2}|\scien{-4.28}{-3} |\scien{-1.14}{-5} \cr
4.0 |$-0.189$     |\scien{-2.44}{-2}|\scien{-1.00}{-2} |\scien{-3.02}{-4} \cr
10.0|$-0.174$     |\scien{-2.69}{-2}|\scien{-1.29}{-2} |\scien{-1.32}{-3} \cr
20.0|$-0.172$     |\scien{-2.73}{-2}|\scien{-1.33}{-2} |\scien{-1.74}{-3} \cr
40.0|$-0.172$     |\scien{-2.74}{-2}|\scien{-1.34}{-2} |\scien{-1.87}{-3}
\endtable
\centerline{\fourteenbf Table 1}
\vfill\eject
%%%
%%% Figures
\input epsf
\frontpagetrue
\epsfxsize=0.9\hsize
\epsfysize=0.9\vsize
\centerline{\epsfbox{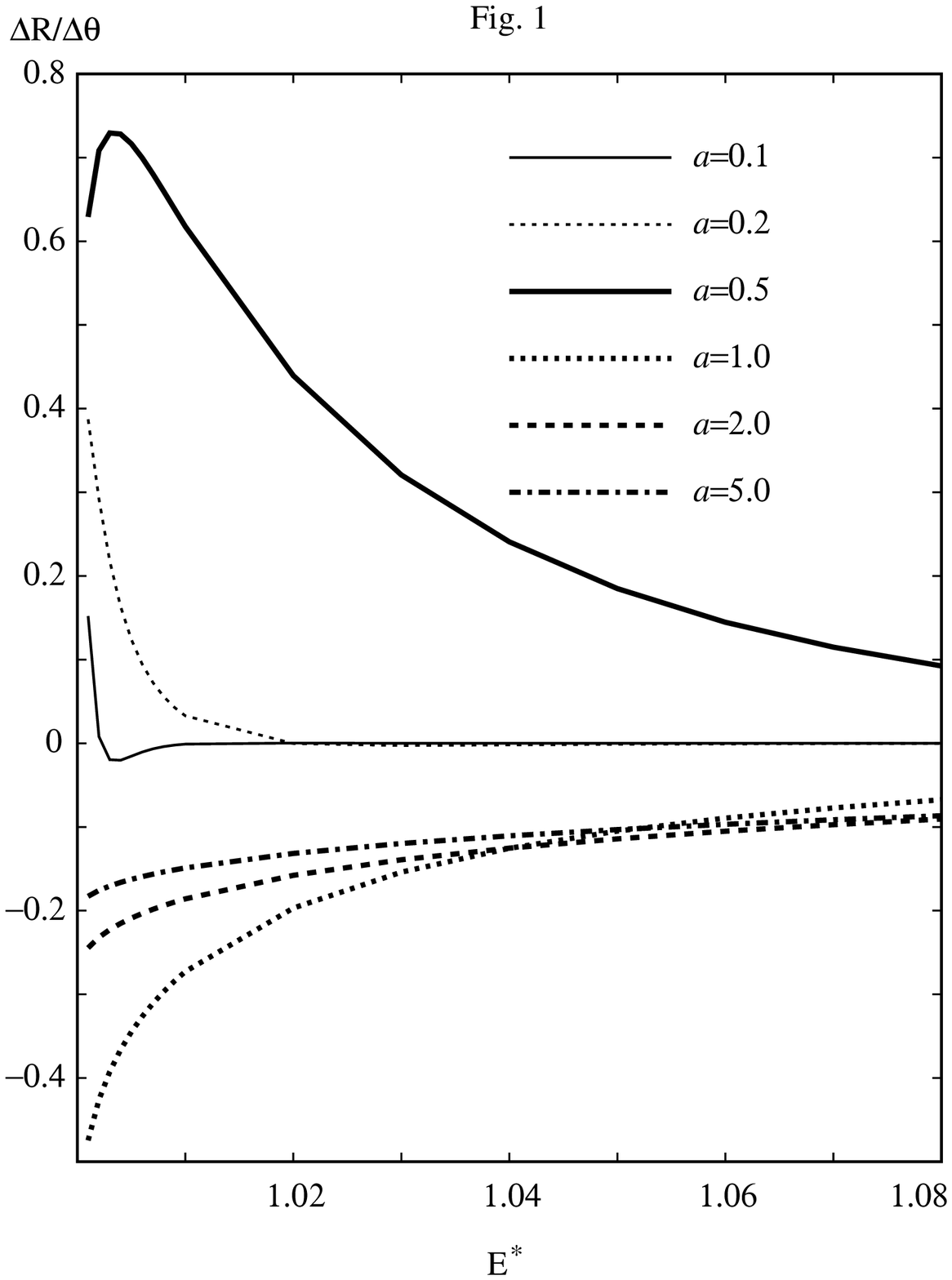}}
\vfill\eject
\frontpagetrue
\epsfxsize=0.9\hsize
\epsfysize=0.9\vsize
\centerline{\epsfbox{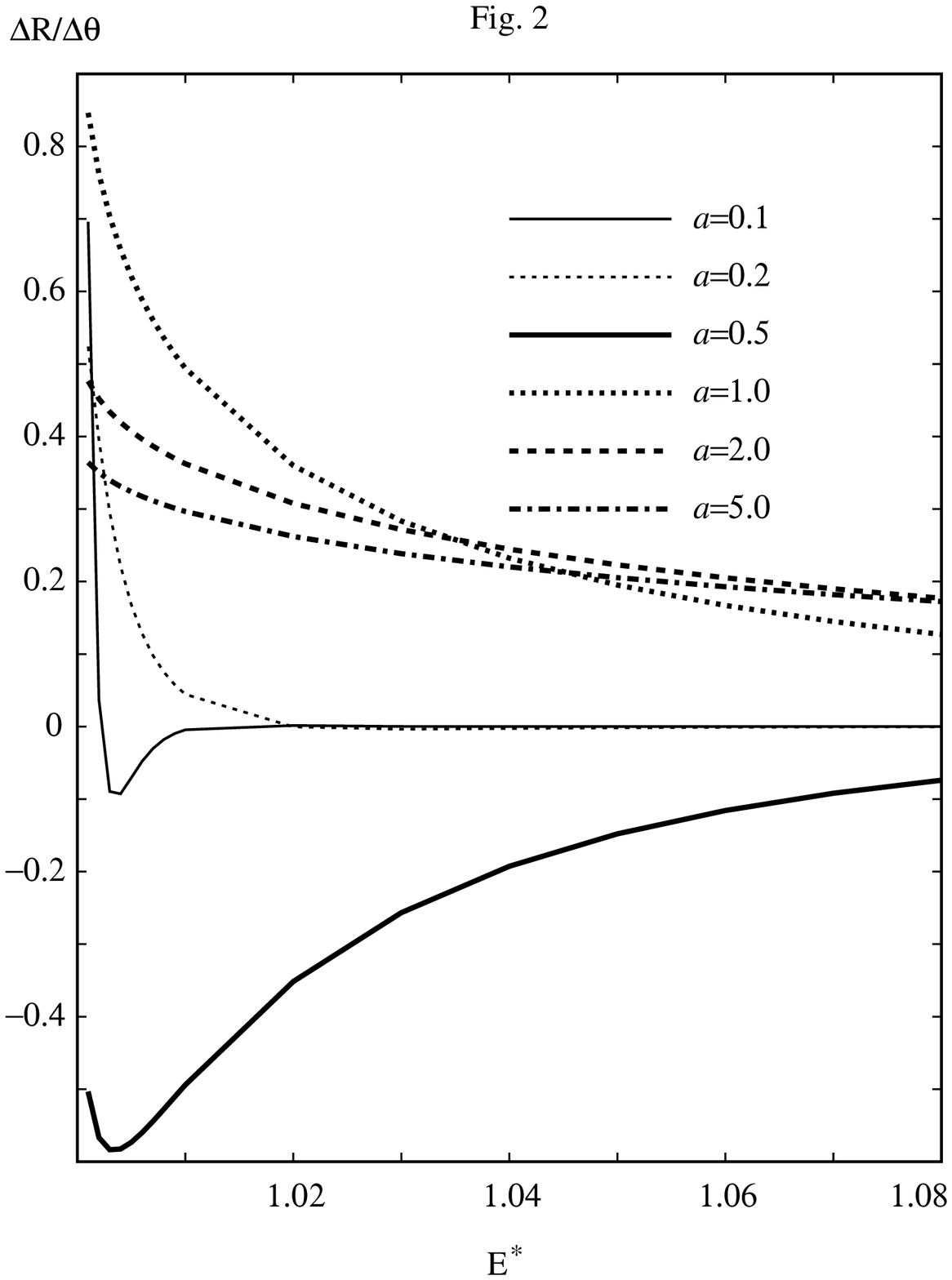}}
%%%
%%%
\bye